\documentclass[pra,twocolumn,superscriptaddress]{revtex4-1}
\usepackage[utf8]{inputenc}
\usepackage{graphicx}
\usepackage{amsmath}
\usepackage{esint}
\usepackage[breaklinks, colorlinks=true, citecolor=blue, urlcolor=blue]{hyperref}
\renewcommand{\section}[1]{{\par\it #1.---}\ignorespaces}
\begin{document}
\title{Suppressed dissipation of a quantum emitter coupled to surface plasmon polaritons}
\author{Chun-Jie Yang}
\affiliation{College of Physics and Materials Science, Henan Normal University, Xinxiang 453007, China}
\author{Jun-Hong An}\email{anjhong@lzu.edu.cn}
\affiliation{School of Physical Science and Technology, Lanzhou University, Lanzhou 730000, China}
\begin{abstract}
Enabling the confinement of light to a scale far below the one of conventional optics, surface plasmon polaritons (SPPs) induced by an electromagnetic field in a metal-dielectric interface supply an ideal system to explore strong quantized light-matter coupling. The fast matter-SPP population exchange reported in previous works makes it a candidate for spin manipulation, but such reversible dynamics asymptotically vanishes accompanying the quantum matter relaxing completely to its ground state. Here, we study the exact dissipative dynamics of a quantum emitter (QE) coupled to SPPs. It is interesting to find that, qualitatively different from conventional findings, the QE can be partially stabilized in its excited state even in the presence of the lossy metal. Our analysis reveals that it is the formation of a QE-SPP bound state which results in such suppressed dissipation. Enriching the decoherence dynamics of the QE in the lossy medium, our result is helpful to understand QE-SPP interactions and apply plasmonic nanostructures in quantum devices.
\end{abstract}

\maketitle
\section{Introduction}
Surface plasmon polaritons (SPPs) triggered by an electromagnetic field along a metal-dielectric interface supply an ideal platform to explore strong quantized light-matter coupling \cite{PhysRevLett.97.053002,cacciola2014ultrastrong,PhysRevLett.108.066401,torma2015strong,PhysRevLett.110.126801}. Such couplings were achieved conventionally in cavity and circuit QED by boosting the interaction times, but at the price of a limited optical bandwidth and diffraction size \cite{wallraff2004strong,niemczyk2010circuit,PhysRevLett.110.043003,RevModPhys.87.1379}. Realizing strong confinement of light below the diffraction limit, SPPs have inspired great interest in studying subwavelength optics and quantum plasmonics \cite{barnes2003surface, tame2013quantum}. Much effort has been made to realize strong or even ultrastrong coupling between SPPs and a quantum emitter (QE) made of, for example, quantum dots, quantum defects, or nitrogen vacancy centers \cite{andersen2011strongly,PhysRevA.87.022312,PhysRevB.87.155417}. It makes surface plasmonics a promising platform to design quantum devices \cite{PhysRevLett.101.190504,berini2012surface,juan2011plasmon,PhysRevLett.108.226803}, where the coupling is a prerequisite. However, the dissipation induced by the lossy metal to the QE severely limits its practical use.

Owing to the strong spatial confinement of the radiation field of the QE near the metal-dielectric interface, the decoherence dynamics of the QE exhibits significant reversibility. Based on the dyadic Green's tensor formalism, the quantization of an electromagnetic field in a lossy medium was developed in Ref. \cite{PhysRevA.53.1818}, from which the quantized interactions between QE and SPPs can be studied. It was found that the spectral density of the SPPs approaches a Lorentzian form when the QE-interface distance is much smaller than the typical wavelength of the SPPs \cite{PhysRevA.82.043845}. In this regime, the decoherence of the QE caused by the SPPs can be approximately stimulated by a Markovian one with QE coupled to an artificial pseudo damping cavity mode \cite{PhysRevB.88.075411,PhysRevLett.117.107401,PhysRevB.93.045422,PhysRevA.94.023818,Varguet:16, PhysRevLett.112.253601,PhysRevB.89.041402,PhysRevB.92.205420}. Thus one can use the well-developed tools in cavity QED to study QE-SPP interactions. Based on this method, the interplay between quenching and strong coupling of QE near a metal nanoparticle \cite{PhysRevLett.112.253601} and the dissipative dynamics of QE near a metal-dielectric interface \cite{PhysRevB.89.041402} were studied, which showed that the QE and the SPPs have a fast population exchange manifesting the strong QE-SPP coupling when the QE-interface distance is small. However, this reversible dynamics tends to vanish in long time limits with the QE asymptotically decaying to its ground state due to the dissipative nature of the metal and the radiative modes to the dielectric. In practical applications of quantum engineering, one always desires that the decoherence of the QE could be stabilized.

\begin{figure}[tbp]
\includegraphics[width=0.9\columnwidth]{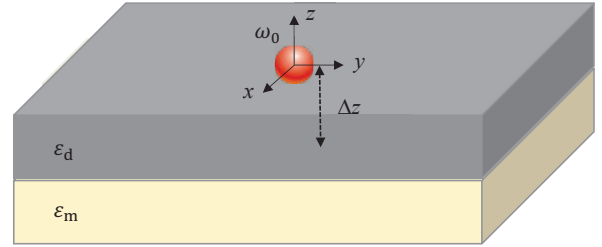}
\caption{A QE with frequency $\omega_0$ placed at a distance $\Delta z$ above a metal-dielectric interface. The QE is embedded in the dielectric adjacent to the metal. $\varepsilon_\text{d}$ and $\varepsilon_\text{m}$ are dielectric functions of the dielectric and the metal, respectively.} \label{Model}
\end{figure}
Inspired by population trapping in structured environments \cite{PhysRevLett.64.2418,Liu2016}, we are interested in how and when the dissipation of a QE coupled to SPPs is suppressed. For this purpose, going beyond the previous cavity-QED treatment, we study the dissipative dynamics of a QE placed at a distance much smaller than the wavelength of the radiation field above a planar metal-dielectric interface. We find that the excited-state population of the QE would be partially preserved in the steady state with a decrease of the QE-interface distance even in the presence of the lossy metal. Our analysis of the physical mechanism reveals that it is the formation of a bound state between the QE and the SPPs which results in this suppressed dissipation. A comparison with the previous cavity-QED treatment \cite{PhysRevB.89.041402} to the QE-SPP coupling indicates that the formation of the bound state would cause the pseudo-cavity-mode treatment breakdown.

\section{System and quantization} \label{system}
We study the interactions between a QE and SPPs induced by the emitted electromagnetic field of a QE in a flat metal-dielectric interface. The system consists of a QE with frequency $\omega_0$ embedded at a distance $\Delta z$ above the metal-dielectric interface (see Fig. \ref{Model}). The metal is characterized by a complex Drude dielectric function $\varepsilon_\text{m} (\omega )=\varepsilon _{\infty }-\omega _{p}^{2}/[\omega (\omega +i\gamma _{p})]$, where $\omega _{p}$ is the bulk plasma frequency, $\varepsilon _{\infty}$ is the high-frequency limit of $\varepsilon_{\text{m}}(\omega)$, and $\gamma _{p}$ is the Ohmic loss of the electromagnetic field in the metal. The distance $\Delta z$ is experimentally controllable \cite{andersen2011strongly}. Here, we choose the dielectric to be germanium with $\varepsilon_\text{d}=25$ and the metal to be silver with $\hbar\omega _{p}=9$ eV, $\varepsilon _{\infty}=5.7$, and $\hbar\gamma _{p}=0.1$ eV in the interested frequency range \cite{PhysRevB.6.4370}. The two media are nonmagnetic and thus the permeabilities are $\mu_\text{d}=\mu_\text{m}\equiv1$.

To describe the quantum features of the QE-SPP interactions, a self-consistent quantization scheme of the electromagnetic fields in the absorbing medium is needed. The scheme was developed based on the dyadic Green's tensor method with the medium's absorption described by a Langevin noise, by which the canonical commutation relations for the field can be guaranteed \cite{PhysRevA.53.1818}. Then the quantized electric field is expressed as
\begin{equation}
\hat{\mathbf{E}}(\mathbf{r},\omega )=\frac{i c^{-2}\omega ^{2}}{\sqrt{\pi \varepsilon_{0}/\hbar}}\int d^{3}\mathbf{r}^\prime\sqrt{\textrm{Im}[\varepsilon_{\text{m}}(\omega)]}\mathbf{G}(\mathbf{r},\mathbf{r}^\prime,\omega )\mathbf{\cdot} \hat{\mathbf{f}}(\mathbf{r}^\prime,\omega ),\label{qtze}
\end{equation}
where $\varepsilon_{0}$ is the vacuum permittivity, $c$ is the speed of light, and $\hat{\mathbf{f}}(\mathbf{r}^\prime,\omega )$ obeying $[\hat{\mathbf{f}}(\mathbf{r},\omega ),\hat{\mathbf{f}}^{\dag}(\mathbf{r}^{\prime },\omega)]=\delta (\mathbf{r}-\mathbf{r}^{\prime })\delta (\omega -\omega ^{\prime })$ are annihilation operators of the electromagnetic field in the absorbing medium. The Green's tensor $\mathbf{G}(\mathbf{r},\mathbf{r}^\prime,\omega)$ denotes the field in frequency $\omega$ at $\mathbf{r}$ triggered by a point source at $\mathbf{r}^{\prime}$ and can be determined by the Maxwell-Helmholtz equation $[{\pmb\nabla}\times{\pmb\nabla}\times-\omega^{2}\varepsilon_\text{m}(\omega)/c^{2}]\mathbf{G}(\mathbf{r},\mathbf{r}^{\prime };\omega )=\mathbf{I}\delta (\mathbf{r}-\mathbf{r}^{\prime })$, where $\mathbf{I}$ is the identity matrix. For general medium geometries, numerical simulations are needed to evaluate $\mathbf{G}(\mathbf{r},\mathbf{r}^{\prime };\omega )$, whereas for symmetric geometries such as spheres \cite{PhysRevA.89.053835}, cylinders \cite{PhysRevB.82.075427}, and planes \cite{PhysRevB.80.155307}, the analytical solutions are attainable. The scheme incorporates the dispersion and the loss of the field in the medium into the Green's tensor. It allows for a complete description of the quantized light-matter interactions by calculating the Green's tensor.

Three distinct modes are triggered by the radiation field of the QE \cite{Pitarke2007}. The first one is the radiative modes propagating into the dielectric. The second one is the damped nonradiative mode absorbed by the metal. The last one is the tightly confined field called SPPs propagating along the metal-dielectric interface. The electromagnetic modes associated with the SPPs enable tight confinement of light on the interface and thus enhance the ultrastrong light-matter interaction.

\section{Exact dynamics} \label{exact-dynamics}
With Eq. \eqref{qtze} at hand, we obtain the Hamiltonian of the QE coupled to its radiation field in the dipole and rotating-wave approximations as
\begin{align}\label{Hmtn}
\begin{split}
\hat{H}=&\hbar\omega _{0}\hat{\sigma}_{+}\hat{\sigma}_{-}+\int d^{3}\mathbf{r}\int d\omega \hbar\omega \mathbf{\hat{f}}^{\dag }(\mathbf{r},\omega
)\cdot \mathbf{\hat{f}}(\mathbf{r},\omega ) \\
&-\int d\omega \lbrack{\pmb\mu}\cdot \hat{\mathbf{E}}(\mathbf{r}_{0},\omega )\hat{\sigma}_{+}+\text{h.c.}],
\end{split}
\end{align}
where $\hat{\sigma}_{+}=|e\rangle \langle g|$ ($\hat{\sigma}_{-}=\hat{\sigma}_{+}^{\dag}$) is the raising (lowering) operator between the ground state $|g\rangle $ and the excited state $|e\rangle $ of the QE. The total excitation number operator $\hat{N}=\hat{\sigma}_{+}\hat{\sigma}_{-}+\int_{0}^{\infty }d\omega\mathbf{\hat{f}}^{\dag }(\mathbf{r},\omega )\cdot \mathbf{\hat{f}}(\mathbf{r},\omega)$ is conserved. Thus when the radiation field is in the vacuum state $|\{0_{\omega}\}\rangle$ initially, the reduced dynamics of the QE is governed by the exact master equation \cite{breuer2002theory}
\begin{equation}
\dot{\rho}(t)=i{\omega(t)\over2}[\rho(t),\hat{\sigma}_{+}\hat{\sigma}_{-}]+{\gamma(t)\over 2}\check{\mathcal{L}}_{\hat{\sigma}_-}\rho(t) ,\label{mastereq}
\end{equation}
where $\check{\mathcal{L}}_{\hat{o}}\cdot=2\hat{o}\cdot\hat{o}^\dag-\{\cdot,\hat{o}^\dag\hat{o}\}$ and $\gamma(t)+i\omega(t)=-2\dot{\alpha}(t)/\alpha(t)$ with $\alpha(t)$ determined by
\begin{equation}
\dot{\alpha}(t)+i\omega _{0}\alpha(t)+\int_{0}^{t}dt^{\prime }K(t-t^{\prime})\alpha(t^{\prime })=0, \label{evolution}
\end{equation}
under the condition $\alpha(0)=1$. Here, $K(\tau)=\int_0^\infty d\omega J(\omega )\exp(-i\omega \tau)$ is the correlation function of the radiation field and $J(\omega )\equiv\omega ^{2}{\pmb\mu} \cdot \textrm{Im}[\mathbf{G}(\mathbf{r}_{0},\mathbf{r}_{0},\omega )]\cdot{\pmb\mu}/(\pi\hbar \varepsilon _{0}c^{2})$ is the spectral density. The action of the absorbing medium on the QE with dipole vector ${\pmb\mu}$ has been incorporated into the spectral density. The localized Green's tensor $\mathbf{G}(\mathbf{r}_{0},\mathbf{r}_{0},\omega )$ implies that the QE interacts with the emitted field at the same position. It is obtained under the dipole approximation, which is valid when the size of the QE is sufficiently small \cite{PhysRevB.86.085304,PhysRevA.93.053803}. All the non-Markovian effects from the radiation field characterized by the convolution in Eq. \eqref{evolution} have been collected into the time-dependent shifted frequency $\omega(t)$ and decay rate $\gamma(t)$ in Eq. \eqref{mastereq}. It can recover the Markovian one under the conditions that the system-field coupling is very weak and the typical time scale of $K(\tau)$ is much smaller than the one of the QE. Then we can obtain the second-order perturbative solution of Eq. \eqref{evolution} as $\alpha(t)=\exp[-(\bar{\gamma}/2 +i\bar{\omega })t]$ with $\bar{\gamma} =2\pi J(\omega _{0})$ and $\bar{\omega }=\omega _{0}+\mathcal{P}\int \frac{J(\omega )}{\omega-\omega _{0}}d\omega $, where $\mathcal{P}$ represents the Cauchy principle value. Then Eq. \eqref{mastereq} reduces to the Born-Markovian approximate one \cite{PhysRevB.82.115334}. To obtain the exact dynamics, a numerical calculation of Eq. \eqref{evolution} is needed.

When the dipole of the QE orients normal to the interface, i.e., ${\pmb\mu}=\mu\mathbf{e}_{z}$, only the component $G_{zz}(\mathbf{r}_{0},\mathbf{r}_{0},\omega )$ contributes to $J(\omega)$. For our layered medium, it can be written in the cylindrical coordinate system as \cite{novotny2012principles}
\begin{equation}
G_{zz}(\mathbf{r}_{0},\mathbf{r}_{0},\omega )=\frac{i}{4\pi k_{\text{d}}^{2}}\int_{0}^{\infty }dk_{\rho }\frac{k_{\rho }^{3}}{k_{z_{\text{d}}}}(1-r_{\text{p}}e^{2ik_{z_{\text{d}}}\Delta z}), \label{Green-ini}
\end{equation}
where $r_{\text{p}}=\frac{\varepsilon _{\text{d}}k_{z_{\text{m}}}-\varepsilon _{\text{m}}(\omega )k_{z_{\text{d}}}}{\varepsilon _{\text{d}}k_{z_{\text{m}}}+\varepsilon _{\text{m}}(\omega )k_{z_{\text{d}}}}$ is the normal Fresnel reflection coefficient with $k_{z_{\text{m}}}$ and $k_{z_{\text{d}}}$ being the axial components of the wave vector in the metal and the dielectric, respectively. Applying the boundary condition $k_{\rho }^2=k_{\text{d}}^{2}-k_{z_{\text{d}}}^{2}= k_{\text{m}}^{2}-k_{z_{\text{m}}}^{2}$, we obtain
\begin{equation}
J(\omega)={3\gamma_0\sqrt{\varepsilon_\text{d}}\omega^3\over 4\pi \omega_0^3}\text{Re}[\int_0^\infty ds{s^3(1-r_\text{p}e^{2ik_{z_\text{d}}\Delta z})\over \sqrt{1-s^2}}],\label{fspec}
\end{equation}
where $s=k_\rho/k_\text{d}$ and $\gamma_0=\omega_0^3\mu^2/(3\pi\hbar\varepsilon_0c^3)$ is the vacuum spontaneous emission rate characterizing the intrinsic
lifetime of the QE.

\begin{figure}[tbp]
\includegraphics[height=.99\columnwidth] {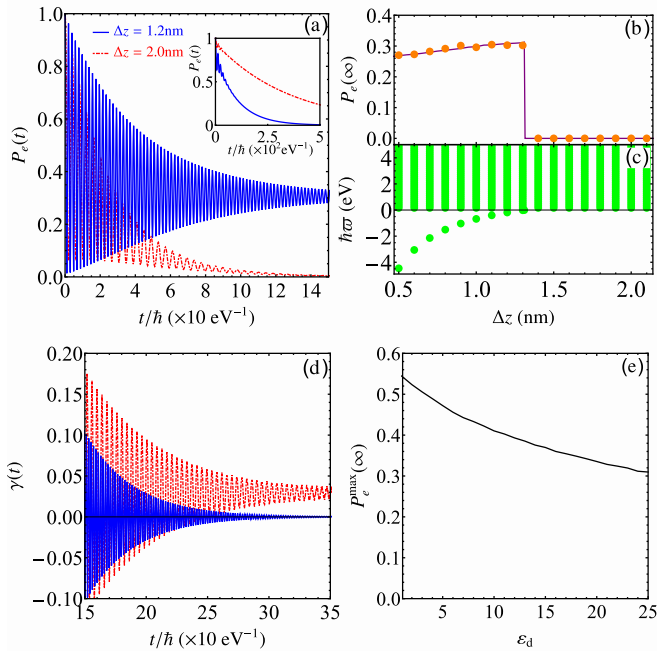}\\
\caption{ (a) Evolution and (b) long-time values of the excited-state population in different $\Delta z$ obtained by numerically solving Eq. \eqref{evolution}. The purple line in (b) shows $Z^2$ evaluated by the bound state. (c) Energy spectrum of the whole system in different $\Delta z$. (d) Decay rate of the QE in different $\Delta z$. (e) Maximal trapped population in different $\varepsilon_\text{d}$. The parameters are chosen as $\hbar\omega_0=1.2$ eV, $\hbar\gamma_0=10^{-4}$ eV, and $\varepsilon_\text{d}=25$. The inset of (a) shows the evolution when $\hbar\gamma_0=10^{-6}$ eV. }
\label{Fig2}
\end{figure}

With the explicit form of Eq. \eqref{fspec}, the exact decoherence dynamics of the QE can be obtained by numerically solving Eq. \eqref{mastereq}. In Fig. \ref{Fig2} (a), we show the evolution of the excited-state population $P_e(t)=\text{Tr}[\hat{\sigma}_+\hat{\sigma}_-\rho(t)]$ of the QE with two characteristic values of $\Delta z$. Under the chosen initial condition $\rho(0)=|e\rangle\langle e|$, one can calculate $P_e(t)=|\alpha(t)|^2$. When the intrinsic decay rate $\gamma_0$ is small, $P_e(t)$ monotonically decays to zero, especially for large $\Delta z$ [see the inset of Fig. \ref{Fig2}(a)]. With increasing $\Delta z$, $P_e(t)$ shows short-time oscillation. When $\gamma_0$ is large, both of the results show that the dynamical oscillation becomes significant. It represents a rapid excitation exchange and thus manifests the strong interactions between the QE and the SPPs. With decreasing $\Delta z$, the dynamics oscillates more rapidly, which is caused by the a stronger near-field confinement of the SPPs and thus a stronger QE-SPP interaction. This quick population exchange makes the hybrid system a promising platform in spin manipulation that demands a fast field response \cite{Peng20151}. A remarkable difference between the two results is their steady-state behaviors. One can see that although $P_e(t)$ approaches zero when $\Delta z=2.0$nm, which is consistent with the previous result \cite{PhysRevB.89.041402}, it tends to a stable non-zero value when $\Delta z=1.2$nm, which is in sharp contrast to the previous result and represents a dissipation suppression of the QE. It indicates that the non-Markovian effect manifests its action on the QE not only in its transient dynamical process, but also in its steady state.

Physically, the decoherence suppression is due to the formation of a bound state between the QE and its environment \cite{PhysRevA.81.052330,PhysRevX.6.021027}. To reveal this, we make a Laplace transform to Eq. \eqref{evolution} and obtain  $\tilde{\alpha}(s)=[s+i\omega_0+\tilde{K}(s)]^{-1}$ with $\tilde{K}(s)=\int_{0}^\infty{J(\omega)\over s+i\omega}d\omega$. According to the Cauchy residue theorem, the inverse Laplace transform can be done by finding the poles of $\tilde{\alpha}(s)$
\begin{equation}
y(\varpi)\equiv\omega_0-\int_{0}^\infty {J(\omega)\over \omega-\varpi}d\omega=\varpi,~(\varpi=is).\label{yee}
\end{equation}
It is interesting to see that the roots of Eq. (\ref{yee}) multiplied by $\hbar$ are just the eigenenergies in the single-excitation subspace of Eq. \eqref{Hmtn} \cite{PhysRevA.81.052330}. It is understandable from the fact that the decoherence of the QE induced by the vacuum environment is governed by the single-excitation process of the whole system. Since $y(\varpi)$ is a monotonically decreasing function when $\varpi<0$, Eq. (\ref{yee}) has one discrete root if $y(0)<0$. It has an infinite number of roots in the region $\varpi>0$, which form a continuous energy band. We name this discrete eigenstate with eigenenergy $\hbar\varpi_\text{b}<0$ the bound state. Its formation would have profound consequences on the decoherence dynamics. To see this, we make the inverse Laplace transform and obtain
\begin{equation}
\alpha(t)=Ze^{-i\varpi_\text{b}t}+\int_{i\epsilon+0}^{ i\epsilon+\infty}{d\varpi\over 2\pi}\tilde{\alpha}(-i\varpi)e^{-i\varpi t},\label{invlpc}
\end{equation}where $Z=[1+\int_{0}^\infty{J(\omega)\over (\varpi_\text{b}-\omega)^2}d\omega]^{-1}$ and the second term contains contributions from the continuous energy band. Oscillating with time in continuously changing frequencies, the second term in Eq. \eqref{invlpc} behaves as a decay and tends to zero due to out-of-phase interference. Therefore, if the bound state is absent, then $\lim_{t\rightarrow \infty}\alpha(t)=0$ characterizes a complete decoherence, while if the bound state is formed, then $\lim_{t\rightarrow \infty}\alpha(t)=Ze^{-i\varpi_\text{b}t}$ implies the dissipation suppression.

Figure \ref{Fig2}(b) shows the steady-state population $P_e(\infty)$ as a function of $\Delta z$ obtained from the exact dynamical evolutions and the bound state. It reveals that $P_e(\infty)$ matches well with $Z^2$ obtained by the bound-state analysis. Figure \ref{Fig2}(c) shows clearly that the regions where the QE population is preserved corresponds exactly to the one where a bound state is formed in the environmental bandgap. Figure \ref{Fig2}(d) is the decay rate $\gamma(t)$ of the QE. The fast oscillations manifest the rapid energy exchange between the QE and the SPPs. It is interesting to see that $\gamma(t)$ asymptotically approaches zero with the formation of the bound state. It indicates that the bound state, as a stationary state of the composite system, has an infinite lifetime. Such behavior implies that the plasmonic nanostructure in the presence of the bound state can be an ideal platform for the coherent manipulation of QE dynamics, even it is inherently dissipative. This is helpful for utilizing plasmonic nanostructures in quantum devices, such as entanglement generators \cite{PhysRevLett.106.020501,1367-2630-15-7-073015}.

It is noted that the parameters used in our calculation are experimentally accessible. The QEs could be quantum dots with $\hbar\gamma_0\simeq 10^{-6}$ eV or J aggregates with $\hbar\gamma_0\simeq 10^{-4}$ eV  \cite{PhysRevB.80.155307,FIDDER1990529}. Although determining the condition when the bound state is formed, the explicit value of $\varepsilon_\text{d}$ does not change our conclusion. We plot in Fig. \ref{Fig2}(e) the maximal trapped population for different $\varepsilon_\text{d}$ by optimizing the distance $\Delta z$. As can be seen, population trapping always exists and the largest population is achievable for small $\varepsilon_\text{d}$. It is meaningful for realizing quantum information processing of the QE via the SPPs.

\section{Breakdown of previous pseudomode description} \label{c-QED}
The spectral density takes a Lorentzian form under the so-called quasistatic approximation \cite{PhysRevA.82.043845}. Assuming $c\rightarrow\infty$ ($k\rightarrow0$), we have $ik_{z_{\text{m}/\text{d}}}=k_{\rho }$. Then the nonradiative part of the Green's function in the second term of Eq. \eqref{Green-ini} can be calculated as $G_{zz}=-\frac{c^{2}}{16\pi \omega ^{2}\varepsilon_{\text{d}}}(\frac{1}{\Delta z})^{3}\frac{\varepsilon_{\text{d}}-\varepsilon _{\text{m}}(\omega )}{\varepsilon_{\text{d}}+\varepsilon _{\text{m}}(\omega )}$. The minor contribution of the radiative modes in the dielectric of the first term of Eq. \eqref{Green-ini} has been ignored. Thus the spectral density can be simplified in a Lorentzian form, $J(\omega )=\gamma _{0}\omega _{p}\frac{3}{16\pi }(\frac{\omega _{c}}{\omega _{p}})^{3}(\frac{c}{\omega _{0}\Delta z})^{3}\frac{\gamma _{p}/2}{(\omega -\omega _{c})^{2}+(\gamma_{p}/2)^{2}}$ with $\omega_{\text{c}}=\omega_{p}/\sqrt{\varepsilon_{\text{d}}+\varepsilon_{\infty}}$ being the cutoff frequency of the SPPs. The approximation is widely used in describing the QE in dissipative metals \cite{2040-8986-16-11-114018,PhysRevB.76.035420,PhysRevB.87.195303}.

\begin{figure}[tbp]
\includegraphics[width=\columnwidth] {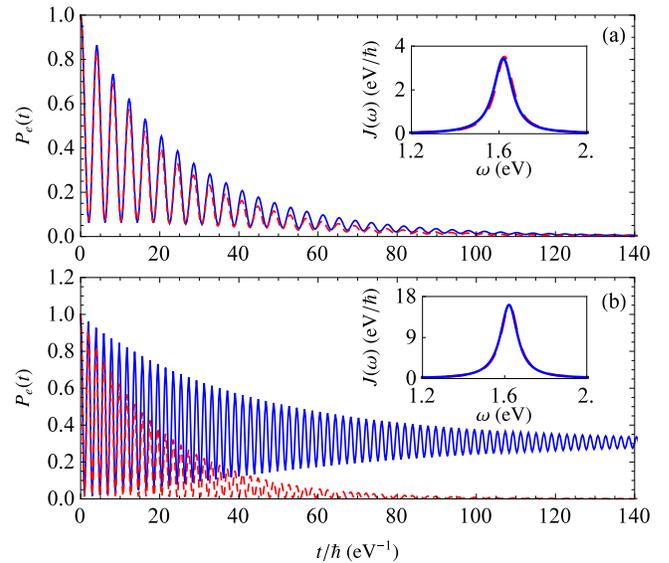}\\
\caption{Evolution of the excited-state population when $\Delta z=2.0$nm in (a) and $1.2$nm in (b) obtained by numerically solving Eq. \eqref{evolution} (blue solid line) and Eq. \eqref{masterdfd} (red dashed line). The insets are spectral the density calculated exactly (blue solid line) and under the quasistatic approximation (red dashed line). Other parameters are the same as Fig. \ref{Fig2}.}
\label{Fig3}
\end{figure}

To ease the difficulty in solving the integro-differential equation \eqref{evolution}, an effective description of the exact dynamics was proposed in Ref. \cite{PhysRevB.89.041402}. With the simplified Lorentzian spectral density, the non-Markovian dynamics of the QE induced by the SPPs was converted into a Markovian one by introducing the pseudocavity mode. Then one can effectively deem that the QE coherently interacts with a pseudocavity mode with frequency $\omega_{\text{c}}$ and damping rate $\gamma_{p}$. Thus the dynamics is governed by
\begin{align}
\begin{split}
\dot{\rho}(t) &=i[\rho(t),\omega_{0}\hat{\sigma}_{+}\hat{\sigma}_{-}+\omega _{\text{c}}\hat{a}^{\dag }\hat{a}+g(\hat{a}\hat{\sigma}^{\dag }+\hat{a}^{\dag }\hat{\sigma})] \\
&+(\gamma _{p}/2)\check{\mathcal{L}}_{\hat{a}}\rho(t), \label{masterdfd}
\end{split}
\end{align}
where $g^{2}=\gamma _{0}\omega _{p}\frac{3}{16\pi }(\frac{\omega _{c}}{\omega _{p}})^{3}(\frac{c}{\omega _{0}\Delta z})^{3}$ is the coupling strength between the QE and the pseudomode. Here, we compare our exact treatment with this effective description. Figure \ref{Fig3} shows the excited-state population evolution of the QE in different $\Delta z$ obtained from numerically solving Eq. \eqref{evolution} and from the effective model \eqref{masterdfd}. The insets plot the spectral density $J(\omega)$ calculated exactly and under the quasistatic approximation. One can see that $J(\omega)$ in both figures shows little deviation from the Lorentzian form, which means that the quasistatic approximation works well. A comparison with the dynamics reveals that the effective model gives an almost perfect description of the decoherence when the bound state is absent [see Fig. \ref{Fig3}(a)], but it shows dramatic differences from the exact dynamical and steady-state results when the bound state is present [see Fig. \ref{Fig3}(b)]. This demonstrates that the effective model fails to capture the correct physics when the QE-SPP bound state is formed.

The breakdown of the pseudomode description can be physically understood as follows. The equivalence between this description of the QE dynamics is established on an analytic continuation of the lower limit of the frequency integral in the correlation function $K(\tau)$ and thus Eq. \eqref{yee} from zero to $-\infty$ \cite{breuer2002theory}. This continuation is mathematically convenient but might miss the physics. The semi-infinity of the environmental energy spectrum is destroyed and no energy gap can support the formation of the bound state in the effective model. Thus the effective model \eqref{masterdfd} would fail to discover the suppressed dissipation when the bound state is actually formed.

\section{Conclusion}\label{con}
We have studied the dissipative dynamics of a QE near a flat metal-dielectric interface supporting SPPs. Due to strong field confinement near the interface, the decoherence dynamics of the QE shows a significant reversibility. It is remarkable to find that the QE may not decay completely to its ground state and part of its excited-state population could be trapped in the steady state. This qualitatively different result from the conventional one reveals that the non-Markovian effect may not only cause a quantitative correction to the short-time dynamics, but also induce a qualitative change to the steady state of the QE. Our analysis demonstrates that such a qualitative change is essentially due to the formation of a bound state of the QE-SPP system. Our result is beneficial for understanding QE-SPP interactions and applying plasmonic nanostructures in quantum devices.

\section{Note added}
Recently, we became aware of a related work \cite{PhysRevB.95.075412}.

\section{Acknowledgments}
This work is supported by the National Natural Science Foundation (Grant No. 11474139), by the Fundamental Research Funds for the Central Universities, and by the Program for New Century Excellent Talents in University of China.

\bibliography{bound-spp}
\end{document}